\documentclass[a4paper,11pt]{article}
\usepackage{epsfig}
\usepackage{graphicx}

\makeatletter

\@addtoreset{equation}{section}
\makeatother
\def \be {\begin{equation}}
\def \ee {\end{equation}}
\def \bea {\begin{eqnarray}}
\def \eea {\end{eqnarray}}
\def \nn {\nonumber}

\def \a {\alpha}
\def \b {\beta}
\def \g {\gamma}

\def \d {\delta}

\def \m {\mu}
\def \n {\nu}
\def \k {\kappa}

\def \s {\sigma}
\def \r {\rho}
\def \o {\omega}

\def \th {\theta}
\def \Th {\Theta}

\def \t {\tau}

\def \p {\partial}
\def\bd{\begin{document}}
\def\ed{\end{document}}
\def\nn{\nonumber}
\def\bea{\begin{eqnarray}}
\def\eea{\end{eqnarray}}
\let\bm=\bibitem
\let\la=\label

\def\N{{\cal N}}
\def\sst{\scriptscriptstyle}
\def\thetabar{\bar\theta}
\def\Tr{{\rm Tr}}
\def\one{\mbox{1 \kern-.59em {\rm l}}}

\def\a{\alpha}      \def\da{{\dot\alpha}}
\def\b{\beta}       \def\db{{\dot\beta}}
\def\c{\gamma}  \def\C{\Gamma}  \def\cdt{\dot\gamma}
\def\d{\delta}  \def\D{\Delta}  \def\ddt{\dot\delta}
\def\e{\epsilon}        \def\vare{\varepsilon}
\def\f{\phi}    \def\F{\Phi}    \def\vvf{\f}
\def\h{\eta}
\def\k{\kappa}
\def\l{\lambda} \def\L{\Lambda}
\def\m{\mu} \def\n{\nu}
\def\o{\omega}
\def\P{\Pi}
\def\r{\rho}
\def\s{\sigma}  \def\S{\Sigma}
\def\t{\tau}
\def\th{\theta} \def\Th{\Theta} \def\vth{\vartheta}
\def\X{\Xeta}
\def\z{\zeta}
\def\w{\wedge}
\def\u{\underline}
\def\hs{\hspace}

\def\cA{{\cal A}} \def\cB{{\cal B}} \def\cC{{\cal C}}
\def\cD{{\cal D}} \def\cE{{\cal E}} \def\cF{{\cal F}}
\def\cG{{\cal G}} \def\cH{{\cal H}} \def\cI{{\cal I}}
\def\cJ{{\cal J}} \def\cK{{\cal K}} \def\cL{{\cal L}}
\def\cM{{\cal M}} \def\cN{{\cal N}} \def\cO{{\cal O}}
\def\cP{{\cal P}} \def\cQ{{\cal Q}} \def\cR{{\cal R}}
\def\cS{{\cal S}} \def\cT{{\cal T}} \def\cU{{\cal U}}
\def\cV{{\cal V}} \def\cW{{\cal W}} \def\cX{{\cal X}}
\def\cY{{\cal Y}} \def\cZ{{\cal Z}}

\def\ua{\underline{\alpha}} \def\ubb{\underline{\beta}}
\def\ug{\underline{\gamma}}
\def\ub{\underline{\phantom{\alpha}}\!\!\!\beta}
\def\uc{\underline{\phantom{\alpha}}\!\!\!\gamma}
\def\um{\underline{\mu}} \def\un{\underline{\nu}}
\def\ud{\underline\delta}
\def\ue{\underline\epsilon}
\def\una{\underline a}\def\unA{\underline A}
\def\unb{\underline b}\def\unB{\underline B}
\def\unc{\underline c}\def\unC{\underline C}
\def\und{\underline d}\def\unD{\underline D}
\def\une{\underline e}\def\unE{\underline E}
\def\unf{\underline{\phantom{e}}\!\!\!\! f}\def\unF{\underline F}
\def\unm{\underline m}\def\unM{\underline M}
\def\unn{\underline n}\def\unN{\underline N}
\def\unp{\underline{\phantom{a}}\!\!\! p}\def\unP{\underline P}
\def\unq{\underline{\phantom{a}}\!\!\! q}
\def\unQ{\underline{\phantom{A}}\!\!\!\! Q}
\def\unH{\underline{H}}
\def\ul{\underline}

\def\As {{A \hspace{-6.4pt} \slash}\;}
\def\bs {{b \hspace{-6.4pt} \slash}\;}
\def\Ds {{D \hspace{-6.4pt} \slash}\;}
\def\ds {{\del \hspace{-6.4pt} \slash}\;}
\def\ss {{\s \hspace{-6.4pt} \slash}\;}
\def\ks {{ k \hspace{-6.4pt} \slash}\;}
\def\ps {{p \hspace{-6.4pt} \slash}\;}
\def\pas {{{p_1} \hspace{-6.4pt} \slash}\;}
\def\pbs {{{p_2} \hspace{-6.4pt} \slash}\;}

\def\Fh{\hat{F}}
\def\Vh{\hat{V}}
\def\Xh{\hat{X}}
\def\ah{\hat{a}}
\def\xh{\hat{x}}
\def\yh{\hat{y}}
\def\ph{\hat{p}}
\def\xih{\hat{\xi}}

\def\psit{\tilde{\psi}}
\def\Psit{\tilde{\Psi}}
\def\tht{\tilde{\th}}

\def\At{\tilde{A}}
\def\Qt{\tilde{Q}}
\def\Rt{\tilde{R}}
\def\Nt{\tilde{N}}

\def\at{\tilde{a}}
\def\st{\tilde{s}}
\def\ft{\tilde{f}}
\def\pt{\tilde{p}}
\def\qt{\tilde{q}}
\def\vt{\tilde{v}}
\def\nt{\tilde{n}}

\def\delb{\bar{\partial}}
\def\bz{\bar{z}}
\def\bD{\bar{D}}
\def\bB{\bar{B}}

\def\bk{{\bf k}}
\def\bl{{\bf l}}
\def\bp{{\bf p}}
\def\bq{{\bf q}}
\def\br{{\bf r}}
\def\bx{{\bf x}}
\def\by{{\bf y}}
\def\bR{{\bf R}}
\def\bV{{\bf V}}

\def\d{\delta}\def\D{\Delta}\def\ddt{\dot\delta}

\def\p{\partial} \def\del{\partial}
\def\xx{\times}
\def\uno{\mbox{1 \kern-.59em {\rm l}}}

\begin{document}
\begin{titlepage}
\vskip1cm
\begin{flushright}
UOSTP {\tt 110301}
\end{flushright}
\vskip1.25cm \centerline{\large \bf Holographic Correlation
Functions for Open Strings and Branes } \vskip1.25cm \centerline{
Dongsu Bak$^a$, Bin Chen$^b$ and  Jun-Bao Wu$^c$ } \vspace{1.25cm}
\centerline{\sl a) Physics Department, University of Seoul, 13
Siripdae, Seoul 130-743 Korea} \vskip0.25cm \centerline{\sl b)
Department of Physics, and State Key Laboratory of Nuclear Physics
and Technology,} \vspace{2mm}\centerline{\sl Center for High Energy
Physics, Peking University, Beijing 100871, P.R. China} \vskip0.25cm
\centerline{\sl c) Institute of High Energy Physics and Theoretical
Physics Center for Science Facilities,} \vspace{2mm}\centerline{\sl
Chinese Academy of Sciences, 100049 Beijing, P.R. China}
\vskip0.25cm \centerline{\tt dsbak@uos.ac.kr, bchen01@pku.edu.cn,
wujb@ihep.ac.cn } \vspace{1.5cm} \centerline{ABSTRACT}
\vspace{0.75cm} \noindent

In this paper, we compute holographically the two-point and three-point functions of giant gravitons with open strings. We consider the maximal giant graviton in $S^5$ and the string configurations corresponding to the ground states of $Z=0$ and $Y=0$ open spin chain, and the spinning string in AdS$_5$
corresponding to the derivative 
type impurities in $Y=0$ or $Z=0$ open spin chain as well. We treat
the D-brane and open string contribution separately and find the
corresponding D3-brane and string configurations in bulk which
connect the composite operators at the AdS$_5$ boundary. We apply a
new prescription to treat the string state contribution and find
agreements for the two-point functions. For the three-point
functions of two giant gravitons with open strings and one certain
half-BPS chiral primary operator, we find that the D-brane
contributions to structure constant are always vanishing and the
open string contribution for the $Y=0$ ground state is in perfect
match with the prediction in the free field limit.

\vspace{1.75cm}
\end{titlepage}

\section{Introduction}

The holographic computation of the correlation functions has been an important subject in AdS/CFT correspondence.
Soon after the proposal of the correspondence\cite{Maldacena:1997re,Gubser:1998bc,Witten:1998qj}, the two-point and three-point functions of the operators in the protected sectors have been computed on both sides\cite{Freedman:1998tz, Chalmers:1998xr,Lee:1998bxa,Arutyunov:1999en}. On the field theory side, the correlation functions were
discussed in the free field limit, while on the string side, they were computed in the supergravity approximation. The perfect match has been found due to the non-renormalization theorem, strongly supporting the correspondence. However, for the operators beyond the protected sectors, the holographic computation usually becomes difficult.

Very recently, there has been renewed interest in the holographic computation of the correlation functions of semiclassical strings. The semiclassical string states correspond to the composite operators with large quantum numbers in the field theory\cite{Gubser:2002tv}. And from the dictionary of the AdS/CFT correspondence, at leading order their semiclassical energies, whose quantum corrections are suppressed by the large quantum number, give the quantum dimensions of the corresponding operators, which are the eigenvalues of the dilatation operator and are calculable with the help of integrability. This relation provides not only a nontrivial check of AdS/CFT correspondence beyond the BPS protected sector, but also leads to a better understanding of integrable structures on both sides of the correspondence. The study of the correlation functions of semi-classical string states
is expected to shed light on the string interaction and integrability in super-Yang-Mills theory at non-planar level.

There is a clear physical picture in holographic computation of the
correlation functions of semiclassical string states. One needs to
insert the string state vertex operators in the string path
integral. At strong coupling, the path integral is dominated by a
saddle point. Therefore one has to solve the equations of motions
with the vertex operators as sources, and find the string
configurations whose ends shrink and approach to the boundary of
AdS$_5$ at the insertion points of the composite operators
corresponding to the string states.  The holographic computation of
the two point functions is relatively easy and has been illustrated
in \cite{Buchbinder:2010gg, Janik,Tsuji:2006zn} . The computation
for the higher point functions becomes much more difficult, as the
string configurations satisfying the appropriate boundary conditions
are hard if not impossible to obtain. Nevertheless, using a strategy
in the holographic study of operator product expansion(OPE) of Wilson line and
Wilson surface
operators\cite{Berenstein:1998ij,Corrado99,Giombi:2006de,Chen:2007zzr},
the three-point and four-point functions of two very massive string
states and one or two light string states have been
investigated\cite{Zarembo:2010rr,
Costa:2010rz,Roiban:2010fe,Buchbinder:2010ek}. For other related
studies, see
\cite{Buchbinder:2010vw,Grossardt:2010xq,Hernandez:2010tg,Ryang:2010bn,Arnaudov:2010kk,Georgiou:2010an,Escobedo:2010xs,Russo:2010bt,Park:2010vs}.

In this article, we would like to study the holographic computation of two-point and three-point functions of the giant gravitons with open strings. The giant gravitons are the D3-banes wrapping trivial cycles without collapsing due to their coupling to the background flux\cite{McGreevy00,Grisaru00,{Hashimoto00}}. As the open string may end on the D-brane, one can consider the system of the giant gravitons with open strings, which correspond to determinant-like operators mapping to a class of open spin chain\cite{Berenstein:2005vf,Berenstein:2005fa}. As D-brane is non-perturbative object, it is not clear what kind of string state vertex operators should be inserted in the string path integral. One
possibility is to consider the boundary conformal field theory, which may count the presence of D-brane. However, for the case at hand, the string world-sheet picture for the giant graviton wrapping a sphere of finite size seems to make no sense. Nevertheless, inspired by the study of the usual semi-classical closed
string state, we expect the similar picture could be carried over, at least for the maximal giant graviton wrapping an $S^3$ in $S^5$. As to the leading order of string coupling $g_s$ and $\alpha'$, the D-brane and open string decouple, we may compute their contributions separately. Namely  we expect that D-brane contribution is from a ``fat'' D3-brane configuration whose ends shrink and approach to the insertion points at the AdS boundary. For the open string sector, the contribution is from the open string configuration connecting the insertion points, ending on the brane and
satisfying appropriate boundary conditions. For the two-point functions, the computation is relatively simple. For the three-point function, the holographic computation is quite difficult. As in other cases, we consider the cases with two massive states corresponding to the giant gravitons with open strings and one gravity state corresponding to the half-BPS chiral primary operator. Due to the presence of the giant gravitons, we have to consider the contribution of D3-branes to the three-point structure constants, besides the one from open string. It is remarkable that the D-brane contribution is vanishing, possibly due to the BPS property of the giant graviton in our case.

There is another novel feature in our treatment. We propose to use the Routhian to handle the string state contribution. For all the cases studied in the literature, our prescription gives the same answer. But for the cases discussed in this paper, we have to use the Routhian. This seems to suggest that our prescription is more useful.

In the next section, we give a brief review of open string configurations ending on the maximal giant graviton branes in $S^5$, which correspond to the open spin chains. In Sec. 3, we discuss the two-point functions of the giant gravitons with open strings, including the cases corresponding to the ground state of $Z=0$ open spin chain, the ground state of $Y=0$ open spin chain and the open string in AdS$_5$ and ending on $Z=0$ brane. In Sec.~4, we
investigate the three-point functions. In the small coupling limit, we compute the three-point functions in the free field limit, and in the strong coupling limit, we holographically compute it in the limit of operator product expansion. We find the perfect match for the structure constant $c^Y_{123}$ involving the ground state operator of the $Y=0$ open spin chain. 
Last section is devoted to some conclusions and discussions.

\section{Open strings ending on  giant graviton branes\label{open}}
In this section we shall be describing giant graviton branes and
open strings ending on the maximal giant graviton branes. We shall
also comment on the ${\cal N}=4$ SYM operators corresponding to the
D-branes plus the open string states. In this note we shall be
mainly concerned about giant graviton branes moving in the  $S^5$
part of the 10d spacetime. The $S^5$
 can be described by three complex embedding coordinates $(X,Y,Z)$ satisfying
 the constraint,
 \be
 |X|^2+|Y|^2+|Z|^2=1
 \ee
 where we set the AdS radius to be unity.
 Using the parametrization,
 \be
 X=\sin\th\,\cos\alpha \, e^{i\phi_1},\hs{3ex}
Y=\sin\th\,\sin\alpha \, e^{i\phi_2}, \hs{3ex} Z=\cos\th\, e^{i\phi}\,,
\ee
one finds that
 the  $S^5$ metric becomes
 \be ds^2_{S^5}=d\th^2+\cos^2\th\,
d\phi^2+\sin^2\th\, d\Omega_3^2\,,
\ee
with
\be
d\Omega_3^2=\cos^2\alpha\, d\phi_1^2+d\alpha^2+\sin^2\alpha\, d\phi_2^2\,.
\ee
Defining $r=\sin \th$, one can alternatively write the $S^5$ metric as
\be
ds^2_{S^5}=\frac{1}{1-r^2}\,dr^2+(1-r^2)\,d\phi^2+r^2\,d\Omega^2_3\,,
\ee
 where the coordinate $r$ is ranged over the interval $[0,\,1]$. The AdS$_5$ part can be described
 by the embedding coordinates $Y_I$ in $R^{2,4}$ satisfying the constraint
 \be
 -Y_0^2-Y_5^2 + Y_1^2 + Y_2^2 + Y_3^2 + Y_4^2=-1\,.
 \ee
The global AdS  coordinates  are related to the embedding coordinates by
the relations
\bea
 Y_5+iY_0= \cosh\rho\, e^{i t_{ads}}\,,\ \ \ \  Y_k =\sinh \rho\,\, \hat{n}_k\,,
\eea where $k=1,2,3,4$ and $\hat{n}_k$ with $\hat{n}_k \hat{n}_k=1$
describes a unit $S^3$.

The above three complex coordinates $(X, Y, Z)$ are dual to the
following three complex linear combinations \be X=\Phi^1+i\Phi^2,
\hs{3ex}Y=\Phi^3+i\Phi^4,\hs{3ex} Z=\Phi^5+i\Phi^6, \ee of the six
real scalar fields $\Phi^i$ in the ${\cal N}=4$ super Yang-Mills
side.

The D3 giant graviton brane\cite{McGreevy00,Grisaru00,Hashimoto00} of our main interest below wraps an $S^3$ part of the $S^5$
and rotates around the remaining directions of the $S^5$ while taking a pointlike
trajectory given by $\rho=0$ in the AdS$_5$ part of the geometry. These D3 branes preserve a half of the supersymmetries. We first consider a maximal sized giant graviton
brane. One example of such brane is given by the trajectory
$Z=0$ and $\rho=0$, which is called as $Z=0$ brane.   This $Z=0$ brane
corresponds to the gauge invariant SYM operator ${\cal O}_{D3}= {\rm det}\, Z$\,, which is  a
half 
 BPS
state.
The dimension of this operator is simply given by its engineering dimension $N$, which is protected against
any quantum corrections. Below we shall also consider the $Y=0$ brane which is related to the $Z=0$ brane by
an appropriate $SO(6)$ rotation and dual to the operator ${\cal O}'_{D3}= {\rm det}\, Y$\,.

The correspondence between open strings ending on the giant graviton
brane and the open spin chain operators in the ${\cal N}=4$ SYM
theory is first considered in Ref.~\cite{HM} and further developed in
Refs.~\cite{correa,ABR}\, (for a recent review, see \cite{Zoubos:2010kh}).
There are two
classes of   open string states of our interest below: one is the
open string ending on the $Z=0$ brane and the other is the open
string ending on   the $Y=0$ brane. We shall choose the open string
vacuum oriented along $Z$ direction. Then the open string ending on
the $Y=0$ brane takes  Neumann  boundary condition on the $Z$ plane.
The open string can also end on the $Z=0$ brane with Dirichlet
boundary condition on the $Z$ plane; an additional localized
boundary degree of freedom is necessary at each boundary of $Z=0$.

In the ${\cal N}=4$ SYM theory side, the $Y=0$ brane open spin chain is represented
by composite operators\cite{HM}
\be
{\cal O}_Y=\epsilon^{j_1\ldots j_{N-1}A}_{i_1\ldots i_{N-1}B}\,
Y^{i_1}_{j_1} \cdots Y^{i_{N-1}}_{j_{N-1}}
(Z \ldots Z \chi_1 Z\ldots Z\chi_2 Z\ldots Z)^B_A,
\label{ybrane}
\ee
where $\chi_1,\chi_2,\ldots$ represent other SYM fields of bulk excitations.
An elementary excitation of a single impurity
is organized by the $SU(1|2)^2$ symmetry. It is clear that the ground state
of the $Y=0$ open spin chain is described by a unique vacuum configuration which
is a ${1\over 4}$ BPS state preserving the  $SU(1|2)^2$ symmetry.
The other is the $Z=0$ brane open spin chain
represented by composite SYM operators\cite{HM},
\bea
{\cal O}_Z = \epsilon^{j_1 \cdots j_{N-1} A}_{i_1 \cdots i_{N-1} B}\,
Z_{j_1}^{i_1} \cdots Z_{j_{N-1}}^{i_{N-1}}
(\chi_{L} Z \cdots Z \chi_1 Z \cdots Z\chi_2 Z \cdots \chi_R)^B_A
\,\,.\label{zbrane}
\eea
An important difference of the $Z=0$ brane case  from that of the $Y=0$ brane
is that the $Z=0$ open  
spin chain is connected to
the giant graviton
through boundary
impurities $\chi_L$ and $\chi_R$.
Each  boundary state is organized by the representation of $SU(2|2)^2$ symmetry.
This elementary magnon involves 16 degenerate states with the energy
spectrum\cite{HM}
\be
E_B = \sqrt{1+ {\lambda\over 4\pi^2}}\,.
\label{eb}
\ee
Since the presence of the boundary impurities is essential, the space of corresponding ground states
is organized by the $256$ degenerate states of $SU(2|2)_L \otimes SU(2|2)_R$. Thus
there are no remaining supersymmetries for the ground state of the $Z=0$ open spin chain.

We now turn to the open string description\cite{Bak,BOR} of the open spin chain dynamics.
For the AdS$_5$ part, we use the ansatz
\bea
&& Y_5+ i Y_0 =\cosh \rho(\sigma)\, e^{i\kappa_s \tau}\nn\\
&& Y_1+i Y_2 = \sinh \rho(\sigma)\, \, e^{i\omega
\,\tau}\label{ansatz1} \eea with $Y_3+ i Y_4=0$. On the other hand,
for the $S^5$ part, we use the ansatz \be \theta=\theta(\sigma)\,, \
\ \ \  \phi= \nu\tau \ee with the $S^3$ coordinates $\alpha$,
$\phi_1$ and $\phi_2$ being taken as constants. The string equations
of motion tell us that \bea\label{eqrt}
&& \rho'' = (\kappa_s^2 -\omega^2)\sinh\rho \, \cosh\rho\,,\nn\\
&& \theta'' = \nu^2 \sin \theta\, \cos\theta\,.
\eea
The conformal gauge constraints lead to the condition
\be
{\rho'}^2 + {\theta'}^2= \kappa_s^2 \cosh^2\rho -\omega^2 \sinh^2\rho -\nu^2 \cos^2\theta\,.
\ee
The first equation of (\ref{eqrt}) may be integrated and leads to
\be
{\rho'}^2 = \kappa_s^2 \cosh^2\rho -\omega^2 \sinh^2\rho -\nu^2 k^2\,
\label{rhoeq}
\ee
and the second equation of (\ref{eqrt}) can be integrated to give
\be
{\theta'}^2= -\nu^2 \cos^2\theta +\nu^2 k^2\,
\label{thetaeq}
\ee
where the integration constant is fixed by the conformal gauge constraint.
Now for the $Z=0$ brane, one has the Dirichlet boundary condition $\rho=0$ and $\theta=\pi/2$ at
$\sigma =0,\,\pi$. The solution to the equations (\ref{rhoeq}) and (\ref{thetaeq}) satisfying
the boundary condition at $\sigma=0$ is given by the Jacobi elliptic function as\cite{BOR}
\be
\cosh\rho = {1\over {\rm dn}(\tilde{\omega}\sigma, {\tilde{k}}^2)}\,,\ \ \ \
\sin\theta ={\rm dn} (\nu\sigma, k^2)
\ee
where we have introduced
\be
{\tilde{\omega}}^2 \equiv \omega^2 -\nu^2 k^2\,, \hs{3ex}
{\tilde{\kappa}_s}^2 \equiv \kappa_s^2 -\nu^2 k^2\,, \hs{3ex} {\tilde{k}}^2={{\tilde{\kappa}_s}^2\over{\tilde{\omega}}^2}\,.
\ee
Note that the periodicity of the Jacobi elliptic function ${\rm dn}(x,k^2)$ is given by
$2 K(k^2)$ where we have introduced the complete elliptic integral of the first kind
$K(k^2)$ 
defined by
\bea
K(k^2)=\int_0^1 dx {1\over \sqrt{1-x^2}\sqrt{1-k^2 x^2}}\,.
\eea
Since we are interested in the open string for $0\ \le\ \sigma \ \le\ \pi$ starting
from and ending on the $Z=0$ brane at $\rho=0$, we have the relations
\be
\tilde{\omega}\pi = 2 K({\tilde{k}}^2)\,, \ \ \ \
\nu  \pi = 2 K({{k}}^2) \,.
\ee
The corresponding solution describes a string ending on the $Z=0$ brane spinning
around the $Z$ plane of $S^5$ and $Y_1+iY_2$ of $AdS_5$ at the same time.
The ground state of the $Z=0$ open spin chain
that does not involve any derivative type letters, e.g. $D^S_+ Z$,
is described by $\rho(\sigma)=0$ which leads to the condition
$\kappa_s= \nu k$ with $\tilde{\kappa}_s=0$. Then this ground state configuration carries
the energy $E$ and the $S^5$ R-charge $J$\cite{Bak},
\bea
E &=& \frac{\sqrt{\lambda}}{2\pi} \kappa_s \int_0^\pi d\s \cosh^2 \rho
 = {\sqrt{\lambda}\over \pi} \,\,{k\, K(k^2) } \label{eopen} \\
J &=& \frac{\sqrt{\lambda}}{2\pi} \nu \int_0^\pi d\s \cos^2 \theta
={\sqrt{\lambda}\over \pi} \left[\, K(k^2)-E(k^2)\, \right]\,, \eea
where \bea
 E(k^2)=\int_0^1 dx {\sqrt{1-k^2 x^2}\over \sqrt{1-x^2}}
\eea
is the complete elliptic integral of the second kind.
It is then straightforward to show that
\be
2 E_B= E-J
= {\sqrt{\lambda}\over \pi} \left(1-{4\over e^2} e^{-{2\pi J\over\sqrt{\lambda} }}+ \cdots\right)\,
\ee
in the long string limit of large $J$.
This agrees with $E_B$ in (\ref{eb})  in the large $\lambda$ and the large $J$ limit.
The $J$ dependent correction in this spectrum stems from the
finite size effect of boundary degrees of freedom of the $Z=0$ brane.

Next we turn to the open string description of the $Y=0$ open spin
chains. At $\sigma=0,\,\,\pi$, one has the Dirichlet boundary
condition $\rho=0$ in the AdS$_5$ side and $\theta'=0$ for the $Z$
plane which is parallel to the worldvolume directions of the $Y=0$
brane. Again using the string equations in (\ref{rhoeq}) and
(\ref{thetaeq}), one may construct various solutions but we shall
here simply present a rather trivial solution, \be \theta=0\,, \ \ \
\ \rho=0\, \ee with $\kappa_s=\nu$ with $k=1$. This describes a
pointlike string  rotating in the equator $r=1$, which corresponds
to the ground state of the $Y=0$ open string carrying\cite{Bak} \be E= J=
{1\over 2}{\sqrt{\lambda}\,\nu}\,. \ee One may see that there is no
finite size correction. Since this vacuum state preserves a quarter
of the sixteen supersymmetries, one has the non-renormalization of
the ground state energy $E_g=E-J=0$ using the argument
Ref.~\cite{Basu:2004nt}. The operator dual to this ground state
${\cal O}_Y^J$ will be given in subsection~\ref{field}.

\section{Two-point functions}

\subsection{The crucial role played by Routhian\label{routhian}}

A prescription on the computations of the two-point functions from
the semi-classical string solution was given in \cite{Janik}. One of
the points stressed there is that we cannot simply evaluate the
on-shell actions of the semiclassical solutions. Instead there are
important corrections from convolution with the string-state
wave-functions. For point particles and strings only rotating in
$S^5$ part of the background, these corrections just change the
action into energy. But for the strings rotating in $AdS$ part as
well, the story is more involved. We need to treat certain zero
modes carefully. And the proposed corrections in this case reads
 \bea \Delta S
=-\int d\tau d\sigma (\vec{\Pi}-\vec{\Pi}_0)\cdot (\dot{\vec{Y}}-
\dot{\vec{Y}_0}) \eea where $\vec{Y}_0$ and $\vec{\Pi}_0$ are
respectively so called zero-modes of the coordinate $\vec{Y}$ and
the momentum $\vec{\Pi}$. It is also stated there that we should use
the embedding coordinates $Y$'s which respect the $SO(2, 4)$
symmetry.

Though the above prescription works well for all cases considered there,
we found that this 
turns out to be problematic for more
general cases. As an example, for the open string solution in
subsection~\ref{2p open}, we do not obtain any sensible result
following the prescription. And the problem is already there for
the closed string counterpart of the open string solution. Thus
we need some generalization of the prescription.

Through trial and error, we now propose to use the Routhian to
handle the string state contribution. Our suggestion is as follows.
Whenever there is a conserved quantity $Q_a$ for the solution
(except the energy) with the corresponding cyclic coordinate $y^a$,
we perform the Legendre transform
\bea L_R= L -\sum_a Q_{a}
{\dot{y}}^a\,, \eea to yield the Routhian. For the further variation
of the Routhian, the conserved charge $Q_{a}$ should be held fixed.

For the strings only spinning inside $S^5$, the conserved charges
are three spins $J_1, J_2, J_3$ in $S^5$ and the Routhian coincides
with the energy. Our new prescription goes back to the old one in
\cite{Janik} for these strings. It is not hard to see that it is
also the case for the string rotating in both $AdS_5$ and $S^5$
studied there. However as we have already mentioned, the
generalization of the prescription in \cite{Janik} is needed for
more complicated cases.

We also find that, interestingly, the Routhian plays a similar role when
we compute the two-point function from a giant graviton (a D3 brane
inside $S^5$) in the next subsection.

\subsection{Maximal Giant Gravitons\label{2p brane}}

In this subsection, we will compute the two-point function of local
operators dual to maximal giant graviton which is a D3 moving in
$S^5$. More precisely, the operator we focus on is ${\cal
O}_{D3}={\rm det}\,Z$. As we have reviewed, the dual  object in the gravity side is
the $Z=0$ giant graviton brane.


For this computation, we  have to use the Poincare coordinate
for $AdS_5$. The corresponding metric is given by \be
ds_{AdS_5}^2=\frac{1}{z^2}\Big[-dt^2+\sum_{i=1}^3 dx_i^2+dz^2
\Big]\,. \label{metric1} \ee The two-point function we want to
compute in this subsection is \be \langle {\cal
O}^\dagger_{D3}(t=t_f, {\bf x}={\bf 0})\, {\cal O}_{D3}(t=0, {\bf
x}={\bf 0}))\rangle\,,\ee and, as in \cite{Janik}, we need to first
find the suitable classical D3 brane solution.
 The worldvolume coordinates of this D3 brane is
$(\tau, \theta_1, \theta_2, \theta_3)$, where $\theta_i$ are three
directions $(\alpha,\phi_1,\phi_2)$ of the $S^3$ inside $S^5$. The part of
nontrivial embedding involves
$t(\tau), 
z(\tau), \phi(\tau)$ directions
with $r$  being a constant\footnote{We will narrow down to
the case with $r=1$ shortly.}. The boundary conditions read \be
\big(\,t(-s/2), z(-s/2)\,\big)=(0, \epsilon),\hs{3ex} \big(\,t(s/2), z(s/2)\,\big)=(t_f,
\epsilon). \ee
The resulting induced metric  becomes
\be ds_{ind}^2=\Big[\,\frac{-\dot{t}^2+\dot{z}^2}{z^2}+(1-r^2)\dot{\phi}^2\,\Big] d\tau^2+r^2d\Omega_3^2
\ee
where dot denotes a derivative with respect to $\tau$.

The action of D3-brane is given by \be
I_{D3}=I_{DBI}+I_{CS}=-T_{D3}\int\sqrt{-
\gamma}+T_{D3}\int P[C_4],\ee
where $\gamma_{ab}$ is the induced metric of D3 brane, $P[C_4]$ is the pullback of
Ramond-Ramond four-form potential and $T_{D3}$ is the tension of
D3-brane, \be
T_{D3}=\frac{1}{(2\pi)^3g_s\alpha^{\prime2}}=\frac{N}{2\pi^2}.\ee

For the solution at hand, after integration over $S^3$, the DBI part
of the Lagrangian becomes \be {\cal
L}_{DBI}=-T_{D3}\Omega_3r^3\sqrt{\frac{\dot{t}^2-\dot{z}^2}{z^2}-(1-r^2)\dot{\phi}^2}
,\ee
where $\Omega_3=2\pi^2$ is the volume of unit $S^3$. The
Chern-Simons part is the same as the one in \cite{McGreevy00}: \be
{\cal L}_{CS}=\dot{\phi}Nr^4,\ee
with
\be {\cal L}={\cal L}_{DBI}+{\cal
L}_{CS}.\ee

The angular momentum
carried by the giant graviton brane is given by
 \be J=\frac{\p {\cal L}}{\p
\dot{\phi}}=\frac{Nr^3(1-r^2)\dot{\phi}}{\sqrt{\frac{\dot{t}^2-\dot{z}^2}{z^2}-(1-r^2)\dot{\phi}^2}}+Nr^4.
\label{angular1} \ee And
one has
$J=N$ for the maximal case of $r=1$.

It can be checked that
 \be  t= R\tan\kappa\tau+t_0, \hs{3ex}z=\frac{R}{\cos\kappa\tau},
\label{sol} \ee is the solution of the equations of motion.
The boundary condition 
$z(\pm s/2)=
\epsilon$ implies
\be \frac{ R}{\cos\frac{\kappa s}2}=\epsilon.\ee Since
$\epsilon$ is very small, we can see that $s$ cannot be real. We can
write $s=-i\tilde s$ with $\tilde s$ being real. Then the worldsheet
coordinate $\tau$ and the spacetime coordinate $t$ have to be purely
imaginary. From $t(-s/2)=0, t(s/2)=t_f$, we get \be \kappa\sim
\frac2{\tilde s}\log\frac{|t_f|}{\epsilon}.\ee

As we discussed in the previous subsection in detail, we proposed
that it is the Routhian which gives the correlation function. Now
the only conserved charge besides the energy is the angular momentum
$J$. It is easy to see that for maximal giant graviton, the Routhian
is nothing but the DBI part of the action: \be {\cal
L}_{DBI}=-N\kappa=-\frac{2N}{\tilde s}\log\frac{|t_f|}{\epsilon} \ee
and \be I_{DBI}=\int_{+i{\tilde s}/2}^{-i{\tilde s}/2}{\cal
L}_{DBI}d\tau=-2iN\log\frac{|t_f|}{\epsilon}.\ee We are thus led to
\be\exp(iI_{DBI})=\left(\frac{|t_f|}{\epsilon}\right)^{-2N}\ee which
is the expected result of the two-point function $\langle {\cal
O}_{D3}^\dagger (t_f)\, {\cal O}_{D3}(0) \rangle$.


\subsection{Nonmaximal Giant Gravitons in $S^5$}

We  begin with the D3 brane action \be {\cal L} = -{T_{D3} \over 2
} \sqrt{-\gamma}\left( \gamma^{ab}\partial_a X^m \partial_b
X^n G_{mn}(X) -2\right) +{\cal L}_{CS} \ee where $\gamma_{ab}$
is the world volume metric which is dynamical now. Since ${\cal
L}_{CS}$ is metric independent, the equations for the worldvolume
metric $\gamma_{ab}$ become
\be \gamma_{ab} \left(-1+{1\over
2} \gamma^{cd} \partial_c X^m \partial_d X^n
G_{mn}(X) \right) = \partial_a X^m \partial_b X^n G_{mn}(X)
\,. \ee Taking trace of this equation, one finds \be \gamma^{cd
} \partial_c X^m \partial_d X^n G_{mn}(X) =4\,. \ee
We partially solve the problem by the choice $\gamma_{ij}d\sigma^i
d\sigma^j= d\Omega_3^2$ and integrate the Lagrangian density over
the  $S^3$. The resulting Lagrangian then becomes \be L =-{N\over 2}
r^3 \,{b} \,\left( -\left(  {-{\dot{t}}^2 +{\dot{z}}^2 \over z^2} +
(1-r^2){\dot{\phi}}^2 \right){1\over {b}^2} +1 \right)  + N r^4
\dot{\phi} \ee
where $
b=\sqrt{-\gamma_{00}}$. We have set $\dot{r}=0$
consistently since we are interested in the solution with fixed
radius. The angular momentum \be J=Nr^3(1-r^2){\dot{\phi}
\over b
}+Nr^4, \label{angular2} \ee
is a conserved quantity as a result of
the equation of motion. By the Legendre transform, one now introduce
the Routhian \be L_R =L-\dot{\phi} J ={N\over 2} r^3  \,\left(
{-{\dot{t}}^2 +{\dot{z}}^2 \over z^2\, b 
} - b 
\left( {(J/N
-r^4)^2\over (1-r^2) r^6} +1 \right) \right) \ee One may check that
all the resulting equations of motion are equivalent to the ones got from
the Lagrangian before the transformation.  The equation for $b 
$ reads
\be
 -{-{\dot{t}}^2 +{\dot{z}}^2 \over z^2\, b^2 
 } -  \left(
{(J/N -r^4)^2\over (1-r^2) r^6} +1 \right)=0\,. \ee
Using this,  the
 equation for $r$ becomes \be 6 r^2 \left( {(J/N -r^4)^2\over (1-r^2)
r^6} +1 \right) + r^3 {d\over dr}\left({(J/N -r^4)^2\over (1-r^2)
r^6}\right)=0\,, \ee whose stable solution is $r^2=J/N$. Inserting
this solution to $L_R$, one gets the effective Lagrangian for the
AdS part: \be L_R ={N\over 2} (J/N)^{3\over 2}  \,\left(
{-{\dot{t}}^2 +{\dot{z}}^2 \over z^2\, b 
} - b 
N/J \right)\,. \ee
The desired solution of this system is given by  (\ref{sol})
with $b
=\kappa \sqrt{J/
N}$,
which leads to 
\be L_R = -J\kappa\,. \ee
The
treatment afterward is the same as the maximal giant case.

\subsection{The two-point functions from giant gravitons with open
strings \label{2p open}}

Let us consider the
two-point function  $\langle {\cal O}^{\dagger J}_{Z}(x)\, {\cal O}^J_{Z}(0)\rangle$ with the operator ${\cal
O}_Z$ corresponding to the ground state of $Z=0$ open spin
chain without involving any derivative type operators $D^M$. The  open string
solution dual to the single operator is give in section \ref{open}.
We can use the idea in \cite{Janik} to find the open string solution
for two-point function: \bea && x(\tau)=R \tanh\kappa\tau+x_0,\
z(\tau)=\frac{R}{\cosh\kappa\tau}, \nn\\
&& \sin\theta(\sigma)={\rm dn}(\nu\sigma, k^2)\,, \ \ \ \phi=\nu\tau\,,
\label{zsolution}
\eea
with $i\kappa\equiv\kappa_s=\nu k$.
The boundary
conditions $x(-s/2)=0,\  x(s/2)=x, \ z(\pm s/2)=\epsilon$ give the
following constraints: \be x_0=x/2,\ x=2R\tanh\frac{\kappa s}{2},\
\frac{R}{\epsilon}=\cosh \frac{\kappa s}{2}.  \ee

There are two parts of contributions to this two-point function.
The first part is from the Routhian of the D3-brane which is
discussed in subsection \ref{2p brane}. The second part is from the
Routhian of the open string which is just the same as the energy of
the open string. The computations are the same as the ones for the
closed string case in \cite{Janik}. And finally we obtain \be\langle
{\cal O}^{\dagger J}_{Z}(x) {\cal
O}^J_{Z}(0)\rangle=|x|^{-2N-2E_{open}},\ee with $E_{open}$ given in
(\ref{eopen}). The computation of the two-point function $\langle
{\cal O}^{\dagger J}_{Y}(x)\, {\cal O}^J_{Y}(0)\rangle$ from the
point-like open string solution in section \ref{open} can be treated in a  similar manner.
The string solution is
 \bea && x(\tau)=R \tanh\kappa\tau+x_0,\
z(\tau)=\frac{R}{\cosh\kappa\tau}, \nn\\
&&\theta=0\,, \ \ \phi=\nu\tau\,,\ \ i\kappa=\nu,
\label{ysolution}
\eea together with similar constraints from the boundary
conditions.

Now we turn to the open string inside $AdS_5$ \cite{BOR}.
We try the same ansatz as in (\ref{ansatz1}) \bea
&& Y_5+ i Y_0 =\cosh \rho(\sigma)\, e^{i\kappa_s \tau}\nn\\
&& Y_1+i Y_2 = \sinh \rho(\sigma)\, \, e^{i\omega
\,\tau}, \label{ansatz} \eea 
with $Y_3= Y_4 =0$ to describe the spinning open-string solution
ending on the giant graviton branes. We choose the range of string
coordinate $\sigma$ to be $0\le \sigma \le \pi$. For the $S^5$ part,
we assume that the string stays at a point in $S^5$ that satisfies
the required open-string boundary conditions. The contribution from
this part vanishes in this case.

The equation of motion for the string becomes \bea \rho''
=(\kappa_s^2-\omega^2) \sinh \rho \cosh \rho\,. \eea Integration of
this equation leads to \bea {\rho'}^2 =\kappa_s^2 \cosh^2
\rho-\omega^2 \sinh^2 \rho + a\,, \label{maineq} \eea where $a$ is
the integration constant. We also have the conformal gauge
constraint \bea
&& \dot{\vec{Y}} \cdot \dot{\vec{Y}}+ {\vec{Y}'} \cdot {\vec{Y}'}=0\\
&& \dot{\vec{Y}} \cdot \vec{Y}'=0. \eea The latter condition is
satisfied automatically for any ansatz of the form in (\ref{ansatz})
and the former one is solved by setting $a=0$. Below for the later
purpose, we shall relax this condition $a=0$ and solve the equation
of motion for general $a$.

The equation (\ref{maineq}) is solved by \bea \cosh \rho(\sigma)=
{1\over {\rm dn}(\bar{\omega}\sigma, k^2)}\,, \label{spinsol} \eea
where we have defined \bea \bar{k}^2\equiv {{\bar{\kappa}}_s^2 \over
{\bar{\omega}}^2}\,, \ \ {\bar{\omega}}^2 \equiv \omega^2 +a\,,
\ \ {\bar{\kappa}}_s^2 \equiv \kappa_s^2 +a \,.\eea
Since we are considering the open string starting
from and ending on the giant graviton at $\rho=0$, we find the
relation \bea \bar{\omega} \pi = 2 {K}(\bar{k}^2)\,. \label{omega}
\eea

\begin{figure}[ht!]
\centering  
\includegraphics[width=4cm]{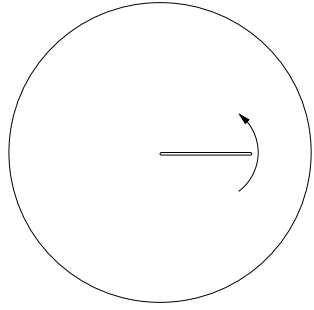}
\caption{\small The open string starting from and ending on the
giant graviton at $\rho=0$ is rotating in the $(Y_1, Y_2)$ plane. }
\end{figure}

The energy and the angular momentum are given by \bea
&&E={\sqrt{\lambda}\over 2\pi} \kappa_s \int_0^{\pi} d\sigma \cosh^2 \rho,\nn\\
&&S={\sqrt{\lambda}\over 2\pi} \omega \int_0^{\pi} d\sigma \sinh^2
\rho \label{ang} \eea which are related to each other by \bea E=
{\kappa_s\over 2}\sqrt{ \lambda}+ {\kappa_s\over \omega} S\,. \eea
Using the solution in (\ref{spinsol}), one finds \bea
&&E={\sqrt{\lambda}\over \pi} {\kappa_s \over \bar{\omega}} {{ E}(\bar{k}^2)\over 1-\bar{k}^2}\,,\nn\\
&& S={\sqrt{\lambda}\over \pi} {\omega \over \bar{\omega}} \Bigl(
{{ E}(\bar{k}^2)\over 1-\bar{k}^2} -{K}(\bar{k}^2) \Bigr)\,. \label{angular}
\eea

As in \cite{Janik}, we do the transformation $Y_0\to iY_4, Y_4\to
iY_0, \kappa_s\to -i\kappa$ and get \bea
&& Y_4=\cosh \rho (\sigma) \sinh\kappa\tau,\  Y_5=\cosh\rho(\sigma)\cosh\kappa\tau,\nn\\
&& Y_1=\sinh \rho (\sigma) \cos \omega\tau,\  Y_2=\sinh\rho
(\sigma)\sin \omega\tau, \ Y_0=Y_3=0.\label{eqm}\eea
In Poincare
coordinate, this solution is
\bea 
x_1&=& e^{\kappa\tau} \tanh\rho (\sigma)\cos \omega\tau  , \nn\\
x_2&=& e^{\kappa\tau} \tanh \rho(\sigma)\sin \omega\tau , \nn\\
z\,\,&=&\frac{e^{\kappa\tau}}{\cosh\rho(\sigma)}\,, \eea
with $t=x_3=0$.

Now we use the same conformal transformation as the one used in
\cite{Janik} and get
\bea 
x_1&=&\frac{\tanh\rho(\sigma)\cos \omega\tau e^{\kappa\tau}+\frac1R
e^{2\kappa\tau}}
{1+\frac2R \tanh\rho(\sigma)\cos \omega\tau e^{\kappa\tau}+\frac1{R^2}e^{2\kappa\tau}},\nn\\
x_2&=&\frac{\tanh\rho(\sigma)\sin\omega\tau e^{\kappa\tau}}{1+\frac2R \tanh\rho(\sigma)\cos\omega\tau e^{\kappa\tau}+\frac1{R^2}e^{2\kappa\tau}},\nn\\
z\,\,&=&\frac{\frac1{\cosh\rho(\sigma)}e^{\kappa\tau}}{1+\frac2R
\tanh\rho(\sigma)\cos\omega\tau
e^{\kappa\tau}+\frac1{R^2}e^{2\kappa\tau}},\eea
again with $t=x_3=0$.
  Notice that for
fixed $\tau$, the value of $z$ depends on $\sigma$ for the current
case. So we need a slight modification of the boundary conditions
used in \cite{Janik}. We choose the following boundary conditions:
\be x_1(\tau_i, \sigma)=0, \hs{3ex} x_1(\tau_f, \sigma)=x,  \hs{3ex}x_2(\tau_i,
\sigma)=x_2(\tau_f, \sigma)=0,\ee \be
{\rm min}_{\,\,0\le
\sigma \le \pi}\,
z(\tau_i, \sigma)={\rm min}_{\,\,0\le
\sigma \le \pi}z(\tau_f, \sigma)=\epsilon. \ee From these, we get
\be x=R,  \hs{3ex}\kappa=\frac1{s}(2\log R-2\log\epsilon+\log(1-k^2)),\ee
where $s=\tau_f-\tau_i$ and we have used the fact that the minimum
value of $\rm{dn} (\bar\omega\sigma, k^2)$ is given by
$\sqrt{1-k^2}$ for our spinning string. Now we rescale $\epsilon$,
such that
 \be x=R, \hs{3ex}
\kappa=\frac1{s}(2\log R-2\log\epsilon).\ee

We remark here that one can also use other boundary condition by
changing the value of $z$ at $t=t_i,\, t_f$ into the maximal value of
$z$ or the mean value of $z$ or the value of $z$ at any $\sigma$.
These boundary conditions give the same results if we always use
the rescaled $\epsilon$ like above.

As we discussed in subsection~\ref{routhian}, we need to use the
Routhian to compute the two-point functions. For the present
problem, the conserved quantity is the angular momentum $S$ and the
corresponding coordinate $\phi$ is defined by ${Y_1+iY_2}=
 |Y_1+iY_2|e^{i\phi}$ with $\dot\phi=\omega$.
Evaluation of the action can be expressed as \be I={\sqrt{\lambda}\over 4\pi}\int
d\tau d \sigma \left(\kappa^2 \cosh^2\rho +\omega^2 \sinh^2\rho
-{\rho'}^2 \right). \ee This leads to the corrected action of the
form, \bea I_R =I-\int d\tau S \omega={\sqrt{\lambda}\over 4\pi}\int d\tau
d\sigma \left(\kappa^2 \cosh^2\rho -\omega^2 \sinh^2\rho -{\rho'}^2
\right) \eea where we use the expression for the angular momentum in
(\ref{ang}). Using the equation of motion in (\ref{maineq}), the
above expression can be reduced to \bea I_R={\sqrt{\lambda}\over 2\pi} \int d\tau
d\sigma (\kappa^2 \cosh^2\rho -a)= 2 
\left( {\kappa\over
2} \sqrt{\lambda}+{\kappa\over \omega}S -{a\over 4\kappa}\right) \log {R/\epsilon}.
\eea We would then like to minimize the above expression with respect to
the modular parameter $s$ that is proportional to $\kappa={i\kappa_s}$.
One further note that $\bar{\omega}$ will be completely determined
as a function of $\bar{k}$ by (\ref{omega}). Using the definition of $\bar{k}$,
$\bar{\kappa}_s$ is also function of only $\bar{k}$. Using the expression
in (\ref{angular}), we find also that $\omega$ is completely fixed
as a function of $\bar{k}$ when we fix $S$ as a constant. Noting
$a={\bar{\omega}}^2-\omega^2$, one finds that $a$ is also
completely determined as a function of $\bar{k}$. Thus
the extremization
condition
 of $I_R$ with respect to modular parameter $s$ is  equivalent to
that of  extremization of $I_R$ with respect to $k$. As summarized
in the appendix, $d I_R/d\bar{k}^2=0$ gives $a=0$, which is the desired
solution of the Virasoro constraint. One then has \bea e^{i (I^{D3}_R+I_R)}
\propto {1\over (x/\epsilon)^{2\Delta}} \eea with $\Delta=N+
{\k_s\over 2}\sqrt{\lambda}+ {\k_s\over \omega}S\,$ including the contributions from the D3-brane.

\section{Three-point functions}

\subsection{Field theory side \label{field}}
In this subsection, we discuss the computation of three-point
function in  ${\cal N}=4$ SYM theory,
  which will be used for the later comparison
with the semiclassical computation.

We begin our discussion with the correlation functions involving
the vacuum-state operator of the $Y=0$ open spin chain,
\be
{\cal O}^J_Y ={d_N^J}\,\,{1 \over (N-1)!}\,\,\epsilon^{j_1\ldots j_{N-1}A}_{i_1\ldots i_{N-1}B}\,
Y^{i_1}_{j_1} \cdots Y^{i_{N-1}}_{j_{N-1}}
(Z^J)^B_A\,.
\ee
In our convention, 
the normalization factor $d_N^J$  is defined by
 the two-point function
\be
\langle{\cal O}^{\dagger J }_{\,\,Y} (x) \, {\cal O}^{ J}_Y (0) \rangle = {1\over (x^2)^{N+J-1}}\,
\ee
with a unit normalization.
Note that the position space propagator of the fields
$X,Y$ and $Z$ is given by
\be
I(x)= {g^2_{\mbox{\tiny YM}}\over 4\pi^2}\, {1\over x^2}\equiv {s_2 \over x^2}\,,
\ee
in the standard convention of the ${\cal N}=4$ SYM theory.

\begin{figure}[ht!]
\centering  
\includegraphics[width=7cm]{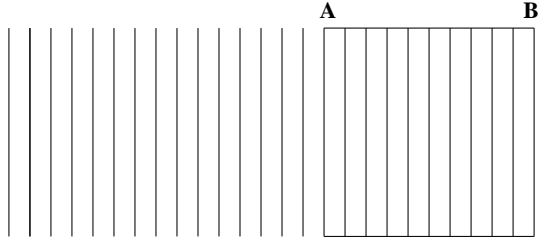}
\caption{\small The relevant Feynman diagram of free field evaluation for the two-point function
 of the $Y=0$ ground state 
 is depicted. We illustrate here 
 the case of
 $N=16$ and $J=11$.
  }
  \label{fig2}
\end{figure}

The relevant Feynman diagram of the free field contractions is depicted in Fig.~\ref{fig2}:
The part involving  separated parallel lines represents the
contractions of elementary fields inside the subdeterminant parts of the operators whereas the remaining
square represents the contractions in the $(Z^J)^A_B$ parts.
With the operator of the normalization
${\cal O}^{J}_Y/d^J_N$, the subdeterminant part has $(N-1)!$ independent ways of contractions
for given indices $A$ and $B$ and the $(Z^J)^A_B$ part involves $N^{J-1}$ factor from the
closed loops of indices. Further including the $N^2$ possibilities of the indices $(A,B)$, one has
 \be
 \langle{\cal O}^{\dagger J }_{\,\,Y} (x) \, {\cal O}^{ J}_Y (0) \rangle\,\, (d^J_N)^{-2}=\big[\,I(x)\,\big]^{N+J-1} N^2\, N^{J-1}\,
 (N-1)!\,.
\ee
Therefore, the normalization
factor is determined as
\be
(d^J_N)^{-2}= (s_2)^{N+J-1}\, N^J\, N!\,.
\ee
This computation is valid for all range of $\lambda$, which can be justified using the
non-renormalization argument of $1/2,1/4$ and $1/8$ BPS operators for the two-point functions in Ref.~\cite{Basu:2004nt}.

Next we turn to the evaluation of the 1/4 BPS three-point correlation function
\be
\langle{\cal O}^{\dagger J+\ell}_Y (x_1) \, {\cal O}^{ J}_Y (x_2) {\cal O}_{Z^{ \ell}} (x_3)\rangle={c^Y_{123}}\,\,{1\over
(x^2_{12})^{N+J-1}\,(x^2_{13})^{\ell}}
\ee
where  ${\cal O}_{Z^{ \ell}}$  denotes  the half BPS chiral primary operator oriented
in the $Z$ direction given by
$
{\cal O}_{Z^{ \ell}} = c_\ell \, {\rm tr}\, Z^{\ell}
$
with $
c_\ell ={(2\pi)^\ell/(\sqrt{\ell} \lambda^{\ell/2}})\,.
$

\begin{figure}[ht!]
\centering  
\includegraphics[width=7cm]{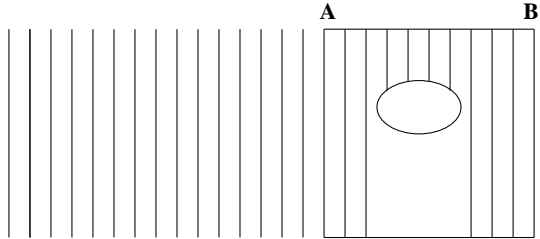}
\caption{\small The relevant Feynman diagram of free field evaluation for $c^Y_{123}$ is depicted here.
We illustrate 
the case of
 $N=16$, $J=7$ and $\ell=4$.
  }
 \label{fig3}
\end{figure} 

The corresponding Feynman diagram for the free field evaluation is illustrated in Fig.~\ref{fig3}.
We note that there are $J-1$ independent ways to put the operator ${\cal O}_{Z^\ell}$ inside the
box and $\ell$ independent ways to pick a particular $Z$ in ${\rm tr} \, Z^\ell$.
The remaining counting of the combinatoric factors is similar to that of the two-point function.
Therefore, one is led to
\be
c^Y_{123}= d^{J+\ell}_N d^{J}_N c_\ell \,N!\, N^{J+\ell-1}\, (s_2)^{N+J+\ell-1}\, (J\!-\!1)\,
\ell={1\over N} \sqrt{\ell}\,(J\!-\!1)\,,
\ee
which is again  valid for all range of the `t Hooft coupling $\lambda$ due to the
nonrenormalization of the 1/4-BPS three-point correlation function\cite{Basu:2004nt}.

This kind of non-renormalization argument does not apply for
the three-point function involving  the operator
${\cal O}_Z^{J}$ which represents a particular ground state of
the $Z=0$ open spin chain carrying  the ($Z$ plane) R-charge $J+N-1$.
 This is because
the three-point 
function of our interest below
\be
\langle{\cal O}^{\dagger J+\ell}_Z (x_1) \, {\cal O}^{ J}_Z (x_2) {\cal O}_{Z^{ \ell}} (x_3)\rangle={c^Z_{123}}\,\,{1\over
(x^2_{12})^{N+J-1+2E_B}\,(x^2_{13})^{\ell}}\,,
\ee
does not preserve  any of the sixteen supersymmetries.
We expect generic perturbative renormalization of the three-point function  and do not have
any field theoretic means to compute
the corresponding structure constant
in the strongly coupled regime.

\subsection{D brane contribution to the structure constants}

In the following two subsections, we shall concern about the
holographic computations of the structure constants $c^Y_{123}$ and
$c^Z_{123}$. This subsection will be devoted to the D-brane
contributions $c_{D3}$. And  we will focus on the open string
contributions $c_{string}$ in the next subsection.
We shall compute the three-point functions in the limit of the
operator product expansion (OPE)
\be R_{12}=|x_1-x_2|\ \ll\
L_{12}\equiv \frac{1}{2}({x_1+x_2})\,,\ee
with the choice of $x_3=0$.
Our prescription of
holographic computations of these three-point functions is along the
way in \cite{Zarembo:2010rr,Costa:2010rz}. The light chiral primary
operator is dual to specific fluctuations of background fields in
string theory side. This fluctuation will lead to the fluctuation
of the classical D3-brane and open string solutions used in the
calculation of the two-point functions. The variant of the on-shell
action of the coupled system of D-brane and open string will give
the three-point functions in the large $N$ and large $\lambda$
limit.
 This can be understood as that we treat the
brane and string as source for the supergravity fields and read off
the OPE coefficients from the coupling to the brane and string of
the bulk supergravity modes dual to the light operators
\cite{Berenstein:1998ij,Giombi:2006de}.

To the leading order of $g_s$ and $\alpha^\prime$, the classical
onshell action of the coupled system of the brane and the string is the sum
of the brane action $I_{D3}$ and the string action $I_{string}$. So
the variation of the on-shell action of the coupled system is equal
to $\delta I_{D3}+\delta I_{string}$. This leads to the fact that
the OPE coefficient $c_{123}$ is the sum of $c_{D3}$ and
$c_{string}$.

The fluctuations of the background $AdS_5\times S^5$  metric
$g_{\m\n}, g_{\a\b}$ corresponds to the half-BPS  chiral primary operators 
are
\cite{Lee:1998bxa} \bea h_{\m\n}&=&(-\frac65\, \ell\,
g_{\m\n}s_I+\frac{4}{\ell+1}\nabla_{(\mu}\nabla_{ \nu)} s_I)\cY^I,\\
h_{\a\b}&=&2\,\ell\, g_{\a\b}\, s_I\,\cY^I, \eea where we use $\mu, \nu, \cdots$
for 
the coordinate indices of the $AdS_5$ part and $\a, \b,
\cdots$  for 
the coordinate indices of the $S^5$ part, and the
round parentheses of $(\mu\nu)$ denotes the symmetric traceless
part. The fluctuation of the background four-form potential is
\cite{Lee:1998bxa}
\bea a_{\m_1\cdots \m_4}&=&-4\epsilon_{\m_1\cdots
\m_5}\nabla^{\m_5}s_I \cY^I,\\
a_{\a_1\cdots\a_4}&=&4\epsilon_{\a_1\cdots \a_5}s_I
\nabla^{\a_5}\cY^I. \eea The spherical harmonic function $\cY^I$ in
the $S^5$ coordinate space is entirely fixed by the choice of the
chiral primary operators: for ${\cal O}_{Z^\ell}=c_\ell{\rm
tr}(Z^\ell)$ and ${\cal O}_{Y^\ell}=c_\ell{\rm tr}(Y^\ell)$,
the corresponding harmonic functions are given by
$2^{-\ell/2}\, Z^\ell$ and $2^{-\ell/2}\, Y^\ell$ respectively.

The linear operation $s^I$ is related to a source
$s_0(\vec{x}^\prime)$ at the boundary of $AdS_5$ through the
following relation: $s^I(\vec{x}, y)=\int d^4\vec{x}^\prime\,
G_\ell(\vec{x}^\prime; \vec{x}, y)s^I_0(\vec{x}^\prime)$ where the
bulk-to-boundary propagator $G_\ell$ is given by \be
G_\ell(\vec{x}^\prime; \vec{x}, y)=g_\ell \left(y\over
y^2+|\vec{x}-\vec{x}^\prime|^2\right)^\ell, \hs{3ex}
g_\ell=\frac{\ell+1}{2^{2-\ell/2}N\sqrt{\ell}}\,. \ee In the OPE
limit $R_{12}\ll L_{12}$, one has \be G_\ell\simeq
g_\ell\,\frac{z^\ell}{L_{12}^{2\ell}}, \ee and also \be
h_{\m\n}\simeq(-2\ell
g_{\m\n}s_I+\frac{4\ell}{z^2}\delta^z_\mu\delta^z_\nu s_I)\cY^I.\ee
The three-point function we want to compute is proportional to
$\frac{\d I_{D3}}{\d s_0(\vec{x}^\prime=0)}$.


We begin with the DBI  action of the D3-brane
\be
I_{DBI}=-T_{D3}\int\sqrt{\gamma}\,.
\ee
Its variation due to the above fluctuation of
the background fields is given by
\bea \d I_{DBI}&=&-\frac12 T_{D3}\int
d^4\sigma \sqrt{ \gamma}\,\gamma^{ab}(h_{\mu\nu}\p_aX^\m
\p_bX^\n+h_{\a\b}\p_aX^\a \p_bX^\b)\nn\\
&=&-\frac12 T_{D3}\int d^4\sigma \sqrt{ \g}\left(-2\ell
s_I\cY^I+\frac{4\ell s_I\cY^I}{\kappa^2z^2}\left(\frac{\p z}{\p
\tilde\tau}\right)^2+6\ell s_I\cY^I\right)\nn \\
&=&-T_{D3}\int d^4\sigma \sqrt{\g}\, 2\,\ell\,
s_I\,\cY^I\left(2-\frac1{\cosh^2\kappa\tilde\tau}\right),\eea where
$\tilde\tau=i\tau$, and we  know that $\tilde\tau$ is real from
subsection~\ref{2p brane}.

Following the prescription just reviewed, replacing $s_I$ by the
bulk-to-boundary propagator, \be G_\ell\simeq
g_\ell\frac{z^\ell}{L_{12}^{2\ell}}=\frac{g_\ell R_{12}^\ell}{2^\ell
L_{12}^{2\ell}\cosh^\ell \kappa\tilde\tau}, \ee one finds that the
contribution of the DBI part of the action to the structure constant
takes the form
\be c_{DBI}=\frac{g_\ell \,\ell N
}{2^{\ell-1}} \int d\tilde\tau\, \kappa\,
\frac{\cY^I}{\cosh^{\ell}\kappa\tilde\tau}\left(-2+\frac1{\cosh^2\kappa\tilde\tau}\right).
\ee

The variation of the Chern-Simons term is given by \be\d
I_{WZ}=\frac{N}{2\pi^2}\int
4\sin^3\theta\cos\theta\sin\alpha\cos\alpha  s^I(-\partial_\theta
\cY^I)\dot{\phi}\,d\tau\wedge d\phi_1\w d\alpha \w d\phi_2\,,\ee and
the contribution from this part to the structure constant is
proportional to
 \be c_{WZ}={
 g_\ell \, N\over 2^{\ell+1}\pi^2 } \,
\int
\frac{4\sin^3\theta\cos\theta\sin\alpha\cos\alpha(-\partial_\theta
\cY^I)\dot{\phi}}{\cosh^{\ell}\kappa\tilde\tau}\, d\tilde\tau\wedge
d\phi_1\w d\alpha \w d\phi_2. \ee

Now we use the above formula to show that the contribution from the
$Z=0$ ($Y=0$) brane part to the structure constants $c^Z_{D3}$
($c^Y_{D3}$) is vanishing. We first consider the case of $c^Z_{D3}$ where
the corresponding spherical harmonic function is
$2^{-\ell/2}\, Z^\ell=2^{-\ell/2}\cos^\ell\th e^{i\ell\phi}$. This  is identically
zero on the $Z=0$ brane which is at $\th=\pi/2$. So the contribution
from the DBI part 
is vanishing. As for the Chern-Simons part, we
notice that $\cos\theta\,\p_\theta (Z^\ell)$ vanishes on the brane for $\ell \ge 1$, and
this leads to the vanishing of the contributions since $\ell \ge 2$ for the chiral primary operators.

The $Y=0$ brane contribution $c^Y_{D3}$ can be related to the the
$Z=0$ brane contribution with the 1/2 BPS operator ${\cal
O}_{Y^\ell}=c_\ell {\rm tr}(Y^\ell)$ by an appropriate $SO(6)$
rotation. The corresponding spherical harmonic function is given by
$2^{-\ell/2} Y^\ell=2^{-\ell/2}\sin^\ell \th \sin^\ell \alpha
e^{i\ell\phi_2}$. The contribution from the DBI part is proportional
to the integral \be c_{DBI} \ \propto\  \int_0^{2\pi}\!\! \! d\phi_2
\,\, e^{i\ell\phi_2}\, \ee which is zero for $\ell \ne 0$. As for
the contribution from the Chern-Simons part, we have $\p_\theta
(Y^\ell)=0$ on the $Z=0$ brane and the corresponding contribution
again vanishes. This shows that $c^Y_{D3}=c^Z_{D3}=0$.


\subsection{Open string contributions}

In this subsection we compute the open string contribution to
the structure constants $c^Y_{123}$ and $c^Z_{123}$. Since we are
dealing with the three-point functions of one light chiral primary
operator and two heavy operators represented by semiclassical string
trajectory, the method developed in
Ref.~\cite{Zarembo:2010rr,Costa:2010rz} is appropriate. The
derivation of the corresponding formula is originally for the closed
string state. The derivation for our open string states is then
essentially the same as the closed string case in
Ref.~\cite{Zarembo:2010rr} but simply replacing the  closed string
trajectory by
  the open string  one. One finds that
\be
c_{string}= {2^\ell (\ell+1)\sqrt{\ell\lambda}\over 8\pi N R^\ell_{12}} \int d^2\sigma{\cal Y}^I z^\ell
\left[
{(\partial \vec{x})^2 -(\partial z)^2\over z^2}- h_{\alpha\beta}\partial X^\alpha \partial X^\beta
\right]
\ee
where one evaluates the above integral with the on-shell open string trajectory in AdS$_5\times S^5$ and
$\ell$ is the dimension of the light chiral primary operator  as before.
The rest is basically straightforward: we use the semiclassical  solution of the open string state
constructed in the previous section to evaluate
the contribution to the structure constant.

Let us now start with the open string contribution to the structure constant $c^Y_{123}$.
Using the open string trajectory in (\ref{ysolution}), one has
\bea
&& {1\over z^2}\Big[
(\partial \vec{x})^2 -(\partial z)^2
\Big]= {2\kappa^2\over \cosh^2 \kappa \tau} -\kappa^2\nn\\
&& h_{\alpha\beta}\,\partial X^\alpha \partial X^\beta =-\kappa^2
\eea
with ${\cal Y}^I=2^{-\ell/2}\cos^\ell\theta e^{i\ell\phi} =2^{-\ell/2} e^{-\ell\kappa\tau}$.
A straightforward computation then leads to the expression,
\be
c^Y_{string}={J\sqrt{\ell}\,(\ell+1)\over  N \, 2^{\ell+1}} \kappa \int^{\infty}_{-\infty}\! d\tau
{e^{-\ell \kappa \tau}\over \cosh^{\ell+2}\kappa\tau} ={1\over N} J\sqrt{\ell}\,.
\ee
Since the $Y=0$ brane contribution is vanishing as computed in the previous section,
our final semiclassical result is given by
\be
c^Y_{123}= c^Y_{D3}+ c^Y_{string}= {1\over N} J\sqrt{\ell}\,,
\ee
which agrees with the field theory result in the large $J$ limit.

Let us now turn to the case of the $Z=0$ open string contribution
to $c^Z_{123}$: By a straightforward evaluation using the solution (\ref{zsolution}), one has
\bea
&& {1\over z^2}\Big[
(\partial \vec{x})^2 -(\partial z)^2
\Big]= {2\kappa^2\over \cosh^2 \kappa \tau} -\kappa^2\nn\\
&& h_{\alpha\beta}\,\partial X^\alpha \partial X^\beta =\kappa^2 -2\kappa^2\, {\rm sn}^2(\nu\sigma, k^2)
\eea
with ${\cal Y}^I=2^{-\ell/2}\, k^\ell\, {\rm sn}^\ell(\nu\sigma, k^2)  e^{-\ell\kappa\tau/k}$.
This leads to  the expression
\bea
&&c^Z_{string}={J(\ell+1)\sqrt{\ell}\over  N \, 2^{\ell+2}} {k^\ell\over K(k^2)\!-\!E(k^2)} \int^{\infty}_{-\infty} dx
{e^{-\ell x/k}\over \cosh^{\ell}x} \nn\\
&&\ \ \ \ \ \ \ \ \  \times \int_0^{2K(k^2)} ds \,\,{\rm sn}^\ell
(s,k^2)\Big( {\rm sn}^2 (s,k^2)-\tanh^2 x \Big) \,, \eea for the
$Z=0$ open string trajectory. Since $k < 1$, the $x$ integral
diverges which implies a divergent $c^Z_{string}=\infty$.
Therefore we have a trouble in this case since our semiclassical
value of $c^Z_{123}$ diverges whereas the field theoretic one should
be finite for a given $\lambda$. We shall comment on this issue in the last section.

\section{Discussions}

In this paper, we studied the holographic computation of two- and
three-point functions of giant graviton with open strings. We
discussed the string configurations corresponding to the ground
states of $Z=0$ and $Y=0$ open spin chain, and the spinning string
in AdS$_5$ corresponding to the derivative 
 type impurities in $Z=0$
open spin chain as well. For the two-point functions, we found both
the D3-brane configuration  and the open string configurations
connecting two insertions of the composite operators at the
boundary. There is only one subtle point: We had to use the Routhian
rather than energy directly in the computation. For the three-point
functions, we consider two giant gravitons with open strings coupled
to a half-BPS chiral primary operator. In the field theory side, we
computed them in the free field limit. In the limit of OPE and with
the configurations for the computation of two-point functions, we
computed the three-point functions holographically. In the
holographic computation, we carefully took the D-brane contribution
into account, which turned out to be vanishing. For the point-like
string dual to the ground state of $Y=0$ open spin chain, we found
perfect agreement. The reason of this agreement may be the
nonrenormalization of this three-point correlation function
\cite{Basu:2004nt} which we mentioned in the field theory
computations. But for the string dual to the $Z=0$ open spin chain,
we obtained a divergent structure constant from holographic
computation. This divergence may be canceled by the $1/N$
corrections to the contributions of D-brane because of the attached
open strings. 
These $1/N$ contributions also include quantum corrections of D branes
and there is currently no available method to compute these quantum
corrections for generic, non-BPS states of D branes.
We also wish that an improved understanding of the prescription for
three-point functions will shed light on this subtle issue.

The computation in this paper is based on the picture that both the D-brane and open string contributes to the computation. This is reasonable as the dual operators include both the part corresponding to the giant graviton and the part corresponding to the open spin chain. However, this picture should be changed for other kinds of open spin chain\cite{Chen:2004mu,DeWolfe:2004zt}. For example, in the case studied in \cite{Chen:2004mu}, there are two kinds of integrable spin chain, closed one and open one, which decouple to the leading order. In other words, the presence of D7-brane just provides boundary conditions for the open semiclassical string. To compute the correlation functions of the open string state, the contribution from D-brane could be neglected safely.

In our work, we discussed the giant gravitons in $S^5$. It would be
nice to consider the case with the giant graviton in AdS$_5$. 
It is also interesting to compute holographically the correlation
functions of these giant gravitons and non-local operators, such as
Wilson loops.

\section*{Acknowledgement}
DB would like to thank Romuald Janik for helpful discussions. DB was
supported in part by NRF SRC-CQUeST-2005-0049409 and  NRF Mid-career Researcher Program 2011-0013228.
BC was in
part supported by NSFC Grant No. 10775002, 10975005. JW would like
to thank Center for High Energy Physics, Peking University and Kavli
Institute for Theoretical Physics China (KITPC) for hospitality at
various stages of this work.

\appendix
\section{Extremal condition for $I_R$}
In this appendix, we shall show that the extremal condition for
$I_R$ with respect to $s$ is given by $a=0$. Using the relation
(\ref{omega}) and the expression in (\ref{angular}),
$\omega$ and ${\bar{\kappa}_s}^2$ are expressed
as \bea
\sqrt{\lambda}\,\omega= {2 S\,{K}(x)\over {{  E}(x)\over 1-x}-{  K}(x)}\,, \ \ \ \
 {\bar{\kappa}_s}^2 =  {4 x \over \pi^2}\,{  K}^2(x) \,,
 \eea where
$x=\bar{k}^2$ and $S$ is taken as a constant in $k$. Then $a$ can be
expressed as a function of $x$ by \bea {\bar{\omega}}^2-\omega^2=a
={4{  K}^2(x)\over \pi^2}\left[ 1- {\pi^2 S^2\over \lambda} \Big({{
E}(x)\over 1-x}-{K(x)}\Big)^{-2} \right] \,.\eea
Note also \bea {da\over
dx}\Big|_{a=0} = {8 \over \pi^2}\,{K}^2(x)\, F(x) \label{da} \eea where
\bea F(x)={d\over dx} \log \left({ {E(x)}\over 1-x}-{  K(x)}\right)
={1\over 2(1-x)}\,\, {{2 {  E(x)}\over 1-x}-{  K(x)}\over {{
E(x)}\over 1-x}-{  K(x)}}\,. \eea We then like to show that \bea
A(x)\equiv{d\over dx} \left({\kappa\over 2}\sqrt{\lambda}+{\kappa\over \omega}S -
{a\over 4\kappa}\right)\Big|_{a=0,\kappa=i\kappa_s}=0\,. \eea This can
be rearranged to \bea i\kappa_s \, A(x)= -{\sqrt{\lambda}\over 4}{d\over dx}
{\bar{\kappa}_s}^2 +\left[ -{\pi\over 4{  K(x)}} {d\over dx}\big(
{\bar{\kappa}_s}^2-a \big)+ x {d\omega\over dx} \right]S\,.
\label{ika} \eea We also note that \bea
{d\over dx}{\bar{\kappa}_s}^2&=&{4\over \pi^2} {{  K(x)}{  E(x)}\over 1-x}\\
\sqrt{\lambda}\,{d\omega\over dx}&=&{d\over dx}\left( 2 {  K(x)} S\over {{
E(x)}\over 1-x} -{  K(x)}\right)- {2 {  K(x)} S\over {{  E(x)}\over
1-x} -{  K(x)}} F(x)\,. \eea Inserting these two together with
(\ref{da}) into $i\kappa_s A(x)$ in (\ref{ika}), one finds that
$i\kappa_s A=0$, which proves that $a=0$ is the extremal condition
for $I_R$.

\end{document}